\documentclass{ws-p8-50x6-00}
\def\pom{{I\!\!P}} 
\begin{document}

\title{The Diffractive Logarithmic Slope \\ and the Saturation
Phenomena}

\author{M.B. Gay Ducati$^{\star}$, V.P. Gon\c{c}alves$^{\dag}$, M.V.T. 
Machado$^{\star}$}

\address{$^{\star}$ Instituto de F\'{\i}sica, Universidade Federal do Rio
Grande do Sul \\  Caixa Postal 15051, 91501-970 Porto Alegre, RS, BRAZIL.\\
E-mail: gay@if.ufrgs.br, magnus@if.ufrgs.br}


\address{$^{\dag}$ Instituto de F\'{\i}sica e Matem\'atica, Universidade
Federal de Pelotas\\
Caixa Postal 354, CEP 96010-090, Pelotas, RS, BRAZIL\\
E-mail: barros@ufpel.tche.br}  


\maketitle

\abstracts{  The logarithmic slope of the diffractive structure function is a
potential observable scanning  the hard and soft contributions in diffraction,
allowing to disentangle the QCD dynamics. We report our calculations
concerning this quantity, in particular the estimates emerging from
the saturation model applied to diffraction dissociation.}

\section{Introduction}

The measurements of the derivative quantity $F_2$-slope\cite{Slopeold} on $Q^2$
 allowed a renewed interest of testing the matching of hard
and soft approaches and provided constraints for the saturation formalisms. The
reported turnover on the $x$ dependence was readly associated with the
transition region between the interplaying domains. The most recent
determinations of the slope\cite{Slopenew} still deserve a better
theoretical description\cite{Thornetalk}, keeping a turnover pattern at fixed
 c.m.s. energy $W$. When we focus in diffractive DIS, in particular
the structure function $F_2^D$, the situation is far from clear: initially
considered as a predominantly soft process, the experimental results suggests
that the diffractive cross section at HERA contains hard and soft
components\cite{diffdata}. However, the hard piece is stronger than the
previously expected. Fortunately, if we consider the pQCD approach,
diffraction stands a more profitable field to study saturation effects than
the inclusive case. This comes from the fact that in DDIS the interaction
probes larger dipole size configuration (soft content) than in the DIS
reaction\cite{GBW}.  Although the quite different pictures considered to
interpret the diffractive measurements, almost all of them fit the data set
properly\cite{modelsroyon}.   We have proposed a derivative  quantity, the
diffractive logarithmic slope, which would help to disentangle the underlying
dynamics in diffractive DIS, settling its validity range, if such observable
is measured. In Ref.\cite{slope1} one perform studies of the proposed
quantity for two sound models, representing the essential features in both
Regge and pQCD formalisms. It was found that important deviations in the
predictions between the models emerge, considering the available kinematic
spectrum in diffractive DIS \cite{blois2001}. Here we report in particular the
calculations considering the saturation pQCD model applied to diffraction
dissociation and its comparison with the non-satured QCD approach, performed
in details in\cite{slope2}. Such results allow to verify the role played  by
the saturation effects in these processes and the extension of the standard QCD
formalism to kinematical regions forbidden perturbatively.

\section{The Diffractive Logarithmic Slope and the Saturation Phenomenon}
\label{satu}

In perturbative QCD, the $\gamma^* p$ process is
described in terms of the photon splitting into a $q\bar{q}$ pair, far
upstream of the nucleon, which then scatters in the proton. This reaction is
mediated by the one gluon exchange which turns out into a multi-gluon one when
the saturation region is approached. A remarkable feature is that the
mechanism leading to the photon dissociation and the further scattering has a
factorizable form, written in terms of a convolution between a photon
wavefunction (dipole transverse size $r$ and quark longitudinal momentum
fraction $z$) and a $q\bar{q}$ cross section. For transverse  and
longitudinally polarized photons the $\gamma^* p$ cross section has the form 
\begin{eqnarray} \sigma_{T,L}(x,\,Q^2) = \int \, d^2  {\bf r} \,\int_0^1 \, d
\alpha \,\, | \Psi_{T,L}(\alpha ,\, {\bf r} )|^2 \,\, \hat{\sigma}(x,\,r^2)
\,\,, \end{eqnarray} \noindent where $\Psi_{T,L}$ are the  photon
wavefunctions. Although they are calculated from perturbation theory, the 
$q\bar{q}$ cross section, $\hat{\sigma}$, has  strong  nonperturbative
contributions that should be modeled. We choose the saturation
model\cite{GBW}, which reproduces the experimental results at both inclusive
and diffractive electroproduction. The dynamics of saturation is present in
the effective dipole cross section in an eikonal way $ \hat{\sigma}(x, \, r^2)
= \sigma_0\,\left[ \, 1- {\rm exp}\left(-\frac{r^2}{4\,R_0^2(x)}\right)
\,\right]$,  where the $x$-dependent saturation scale is  $R_0(x) =
\frac{1}{Q_0} \, \left(\frac{x}{x_0} \right)^{\lambda/2}$.  The normalization
$\sigma_0$ and the remaining parameters were determined from  data with
$x<10^{-2}$.  This model consistently interpolates between the saturation
regime and the scaling regime of $\sigma_T\,(F_2)$.  The elastic scattering of
the  $q\bar{q}$ pair dominates the diffractive $\gamma^* p$ process for not
too large values of the diffractive mass $M_X$. Instead, at large $M_X$ the
emission of a gluon becomes the leading contribution.  Then, the diffractive
cross section is, given the  slope $B_D$, \begin{eqnarray}
\label{slope} \sigma^D(x,Q^2)= \frac{1}{B_D}\frac{1}{16\,\pi}
\,\int \, d^2  {\bf r} \,\int_0^1 \, d \alpha \,\, | \Psi_{T,L}(\alpha ,\,
{\bf r} )|^2 \,\, \hat{\sigma}^2(x,\,r^2)\,.
\end{eqnarray}
In ref.\cite{GBW} it has been found a ratio between the diffractive and the
inclusive cross sections roughly constant, and that the diffractive cross
section is sensitive to the infrared cutoff which is effectively provided by
$2\,R_0(x)$ concluding that diffraction probes the transition region. 
\begin{figure}[t]
\epsfxsize=18pc 
\centerline{\epsfbox{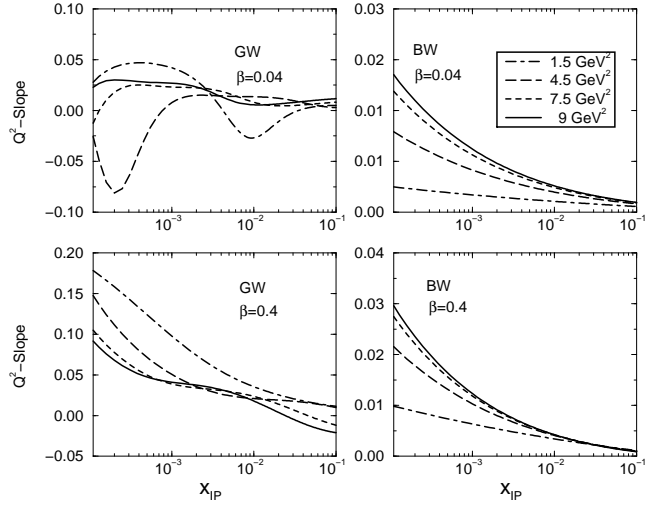}} 
\caption{The $x_{\pom}$ dependence
on the logarithmic slope for the  pQCD model (BW)  and the
the saturation model (GW)
presented at typical $\beta$ values.} 
\label{gwcompxp}
\end{figure}
We performed numerically the $F_2^D$ logarithmic slope, using the same
parameters values from Ref.\cite{GBW}. 
In Fig. (\ref{gwcompxp}), one shows the $x_{\pom}$ behavior for both the approaches
with\cite{GBW} and without saturation\cite{BW}, at typical $\beta$
values. We analyze in particular the transition region between hard and soft
dynamics, settled by the low virtualities $Q^2 \sim 1.5 - 9$ GeV$^2$.  The
saturation model produces a transition between positive and negative slope
values at low $\beta=0.04$, while presents a positive slope for medium and
large $\beta$. This feature is corroborated by the preliminary ZEUS
studies\cite{prelZEUS}. The pQCD approach without saturation\cite{BW}, shows
a positive slope for the whole $Q^2$ and $x_{\pom}$ range. In \cite{slope2},
one observed such a smooth behavior for the pQCD predictions, with a positive
slope in almost the whole range, whereas a sound Regge model predicted
predominantly negative slope values in this kinematical domain, converging to
a flat value for larger $x_{\pom}$. A clear difference between the predictions
of Regge and QCD is the change of signal of the slope with the $Q^2$ evolution
at medium $\beta$. Returning, since  the diffractive cross section is strongly
sensitive to the infrared cutoff, one of the main differences between these
models is the assumption related to the small $Q^2$ region. In the pQCD models
without saturation, an ad hoc cutoff in the transverse momentum is inserted,
as well as the energy dependence of the unintegrated gluon distribution. In
the saturation model, instead,  the saturation radius $R_0(x)$ gives the
infrared cutoff (the saturation momentum scale) and determines the energy
dependence. If this scale is large (1-2 GeV$^2$), then the resulting process
is not soft and can be completely calculated using pQCD methods. Therefore,
the saturation model extends the pQCD approach towards lower $Q^2$ values. We
conclude that, the difference between the behaviors predicted by these two
models for the $x_{\pom}$ spectrum, mainly in the region of small $\beta$ and
medium $Q^2$, is large, which should allow to discriminate the dynamics in
future experimental analyzes.


\end{document}